\documentclass[]{interact}

\usepackage{fix-cm}
\usepackage{type1cm}

\usepackage{amsmath}
\usepackage{booktabs}

\usepackage[hidelinks,hypertexnames=false]{hyperref} 

\usepackage{orcidlink}

\usepackage{graphicx}
\usepackage{float} 
\usepackage{epstopdf}
\usepackage[caption=false]{subfig}

\usepackage[numbers,sort&compress]{natbib}
\bibpunct[, ]{[}{]}{,}{n}{,}{,}
\renewcommand\bibfont{\fontsize{10}{12}\selectfont}

\theoremstyle{plain}

\theoremstyle{definition}

\theoremstyle{remark}

\begin{document}

\articletype{RESEARCH ARTICLE}

\title{Nanosecond Pulsed-Laser Treatment Couples Chloride Removal with Oxide Transformation in Salt-Corroded Carbon Steel}

\author{
\name{Youichi Ishikawa\textsuperscript{a}\,\orcidlink{0009-0003-3735-8233}, 
      Yoshitaka Okuyama\textsuperscript{a}\,\orcidlink{0000-0002-3105-4445} 
      and Daishi Fujita\textsuperscript{a}\,\orcidlink{0000-0001-5726-0645}\thanks{CONTACT Daishi Fujita. Email: dfujita@icems.kyoto-u.ac.jp}}
\affil{\textsuperscript{a}Institute for Integrated Cell-Material Sciences (WPI-iCeMS), Kyoto University, Yoshida, Sakyo-ku, Kyoto 606-8501, Japan}
}

\maketitle

\begin{abstract}
Maintaining carbon steel in marine environments requires surface treatments capable of simultaneously removing corrosion products and chloride contaminants whilst modifying the residual oxide layer. In this study, salt-contaminated SS400 carbon steel was treated using a Q-switched pulsed fibre laser characterized by a full width at half maximum (FWHM) of approximately 150~ns and a decaying tail extending to about 600~ns. The pulse energy was fixed at 1~mJ, and the repetition frequency was varied between 50 and 200~kHz to investigate the effects of cumulative thermal accumulation.
Under 100~W (100~kHz) conditions, WDS-EPMA analysis confirmed that Na and Cl levels dropped to near-background values, demonstrating the comprehensive removal of sea-salt-derived contaminants. SEM observations revealed the transition of the porous rust layer into a dense, laser-modified layer, whilst XPS analysis confirmed the suppression of the $\text{Fe}^{3+}$ satellite feature in the $\text{Fe}~2p$ spectrum, establishing a distinct phase transformation from a haematite-dominant rust to a protective magnetite-rich scale.
These results elucidate a single-step, two-stage mechanism: the high-peak-power leading edge of the pulse drives the ablation of the corrosion layer, whilst the 600~ns trailing tail delivers continuous thermal energy that promotes oxide resolidification and phase transformation. This approach offers a promising, non-contact methodology capable of concurrent decontamination and surface functionalisation in a single processing step.
\end{abstract}


\section{Introduction}

While laser cleaning was initially employed for the conservation and restoration of delicate surfaces \cite{Asmus_OAC}, its scope has since expanded to industrial applications \cite{Steen2010, Zhu2022} aimed at contaminant removal and surface modification. More broadly, pulsed laser processing has emerged as a foundational technology for diverse applications, ranging from nanopatterning and pulsed laser deposition to the precise control of surface chemical states \cite{Bauerle2005}. Early investigations established fundamental mechanisms for particle removal using dry and vapour-assisted methods, leveraging rapid thermal expansion of surfaces and the explosive evaporation of liquid films \cite{Zapka1993, Tam1992}. These theoretical frameworks were subsequently extended to plate-like contaminants by incorporating van der Waals adhesion and laser-induced thermoelastic stress \cite{Song2002}. For carbon steel substrates, the underlying mechanisms involve a complex interplay of thermal expansion, explosive ablation, shock wave generation, transient changes in optical absorptivity, and photothermal energy coupling \cite{Lu1994, Zhou2023}. While these synergistic processes render laser cleaning highly efficient, they demand rigorous parametre control; excessive heat input or unintended microstructural alterations can compromise the structural integrity of the material \cite{Dwivedi2017, Zhou2023_Review, Wu2021}.

For carbon steel infrastructure in marine environments, surface maintenance requires addressing not only visible rust but also chloride ions ($\text{Cl}^-$) retained within rust layers and microscopic pits \cite{Bhandari2015, Melchers2008}. Rust formed in $\text{Cl}^-$-rich atmospheres typically exhibits a stratified oxide structure \cite{Kainuma2009, Alcantara2017}. Specifically, while lepidocrocite ($\gamma\text{-FeOOH}$) commonly constitutes the outer layer, magnetite ($\text{Fe}_3\text{O}_4$) and akaganeite ($\beta\text{-FeOOH}$) tend to develop in the region adjacent to the steel substrate \cite{DelaFuente2016}. Therefore, effective remediation demands a dual strategy encompassing both the removal or consolidation of the porous oxide layer and the elimination of residual salts that could accelerate subsequent corrosion.

Whilst research on continuous-wave (CW) lasers has progressed with the aim of large-area processing \cite{Fujita2020}, pulsed lasers introduce an additional degree of freedom via the temporal structure of energy delivery \cite{Fujita2024_eng, Wu2021}. The pulse width and waveform dictate how the delivered energy is partitioned between the optical absorption depth and the thermal diffusion length \cite{Siano2010}. Broadly, the geometry of the molten zone and subsequent microstructural evolution during laser thermal processing are governed by power density and interaction time \cite{Ayoola2017, Bendoumia2020}. Although low repetition rates are often employed in high-power processing to suppress thermal accumulation \cite{Choubey2014}, intentional heat retention can promote beneficial oxide transformations and surface densification. Crucially, the residual heat supplied by the decaying trailing edge (tail) of a nanosecond pulse can facilitate the reduction of haematite-based rust into a protective magnetite phase. Such surface modifications represent critical factors for corrosion resistance; indeed, oxide layers densified through laser treatment have been demonstrated to significantly enhance the corrosion resistance of iron-based alloys \cite{Ettefagh2020}.

This study investigates the effects of the decaying pulse tail characteristic of Q-switched fibre laser pulses, featuring a full width at half maximum (FWHM) of 150~ns, on chloride removal and oxide transformation in chloride-corroded carbon steel. The degree of thermal accumulation was controlled by fixing the pulse energy at 1~mJ whilst varying the repetition rate between 50 and 200~kHz. This experimental configuration is designed to contrast the ablation driven by the high-intensity leading edge of the pulse against the subsequent residual heating caused by the pulse tail. Ultimately, this approach evaluates the feasibility of simultaneously reducing chloride contamination and transforming the rust layer into a protective surface state primarily composed of magnetite within a single laser operation.

\section{Experimental Methods}
\label{sec:experimental}

\subsection{Material Preparation}
Structural carbon steel plates (JIS SS400; 20~mm $\times$ 20~mm $\times$ 5~mm) were utilised. Prior to the corrosion treatment, each plate was sequentially ultrasonically cleaned in ethanol, acetone, and ethanol to remove organic residues, rinsed with deionised water, and dried under a stream of compressed air. A 3.5~wt.\% NaCl solution was subsequently sprayed onto the cleaned surfaces to simulate an artificial marine atmosphere. The specimens were then exposed to ambient laboratory conditions to promote the formation of a homogeneous red rust layer containing chloride contaminants.

\subsection{Laser Processing System}
\begin{figure}[htb]
\centering
\includegraphics[width=0.9\textwidth]{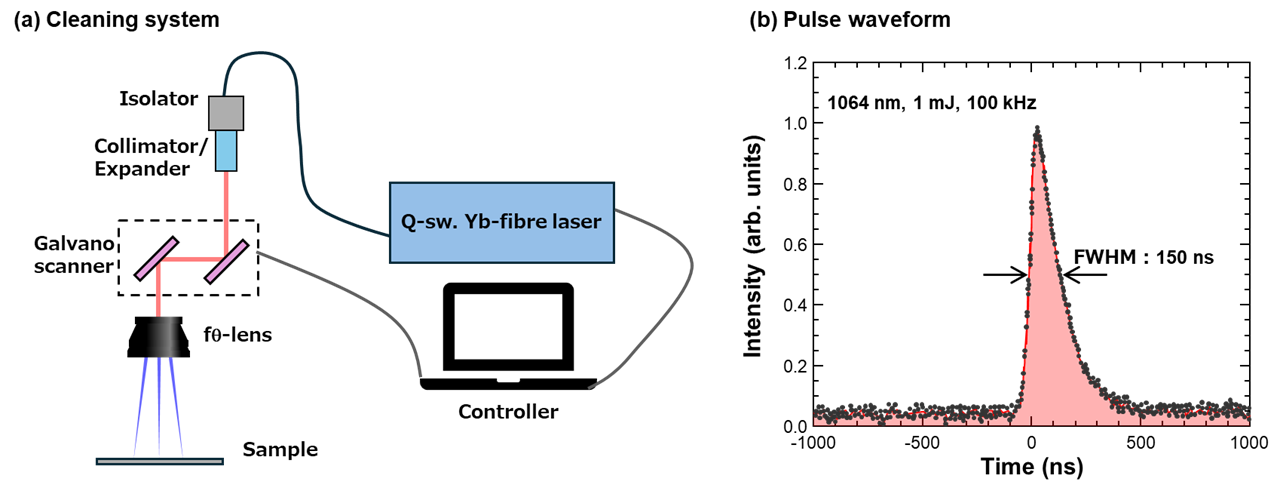}
\caption{(a) Schematic diagram of the laser treatment setup and (b) measured temporal pulse waveform of the Q-switched fibre laser. The waveform exhibits a rapid leading peak followed by a protracted, decaying trailing edge with a full width at half maximum (FWHM) of 150~ns, facilitates cumulative thermal accumulation during repeated irradiation.}
\label{fig:setup_and_pulse}
\end{figure}

Laser treatment was conducted using a Q-switched fibre laser system operated at average powers ($P_{\rm avg}$) ranging from 50 to 200~W. As illustrated in Fig.~\ref{fig:setup_and_pulse}(a), the laser beam was delivered through a galvanometric scanner and focused via an F-theta lens ($f = 254$~mm) to a spot diameter of approximately 80~$\mu$m on the specimen surface. The temporal pulse profile during irradiation was monitored using a high-speed photodetector coupled with a digital oscilloscope. The resultant waveform in Fig.~\ref{fig:setup_and_pulse}(b) exhibits a characteristic asymmetric profile, consisting of a rapid leading peak followed by a protracted, decaying trailing edge with a full width at half maximum (FWHM) of 150~ns.

\subsection{Laser Treatment Conditions}
The repetition rate was varied whilst maintaining a constant pulse energy of 1.0~mJ. To investigate the effects of frequency dependent thermal accumulation on decontamination and surface modification, experiments were conducted at three distinct repetition rates of 50, 100, and 200~kHz (Table~\ref{tab:treatment_params}). At a fixed pulse energy and duration, elevating the repetition rate proportionally increased the average power from 50 to 200~W.

The scanning pitch was configured to minimise spatial overlap between adjacent pulses. The laser beam exhibited a $1/e^{2}$ diameter of 80~$\mu$m, corresponding to a FWHM of approximately 50~$\mu$m. Laser irradiation was performed with a uniform pitch of 80~$\mu$m in both the scanning and hatching directions. This spatial configuration ensured minimal overlap between successive laser spots, thereby isolating the influence of pulse intervals and thermal accumulation on the resultant surface characteristics. To ensure complete removal of the rust layer, the scanning sequence was repeated for five consecutive passes.

\begin{table}
\tbl{Laser treatment parameters investigated across different repetition rates. The scanning pitch was maintained at 80~$\mu$m to position adjacent pulses tangentially, thereby isolating temporal thermal effects.}
{\begingroup
\setlength{\tabcolsep}{15pt} 
\renewcommand{\arraystretch}{1.2} 
\begin{tabular}{lccc}
\toprule
Parameter & Case 1 & Case 2 & Case 3 \\ 
\midrule
Repetition rate (kHz)    & 50    & 100   & 200    \\
Average power (W)        & 50    & 100   & 200    \\
Pulse energy (mJ)        & 1.0   & 1.0   & 1.0    \\
Pulse width, FWHM (ns)   & 150   & 150   & 150    \\
Beam diameter, $1/e^{2}$ ($\mu$m) & 80    & 80    & 80     \\
Beam diameter, FWHM ($\mu$m)       & 50    & 50    & 50     \\
Beam pitch ($\mu$m)      & 80    & 80    & 80     \\ 
\bottomrule
\end{tabular}
\endgroup}
\label{tab:treatment_params}
\end{table}

\subsection{Surface Characterisation}
The treated surfaces were characterised using scanning electron microscopy (SEM), X-ray photoelectron spectroscopy (XPS), and wavelength-dispersive electron probe microanalysis (WDS-EPMA). Surface morphology was observed via a field-emission scanning electron microscope (FE-SEM; Hitachi SU5000). Chemical states within the iron oxide layer were evaluated using an XPS system (ULVAC-PHI PHI 5000 VersaProbe~II) equipped with a monochromatic Al~K$\alpha$ X-ray source, focusing on the Fe~$2p$ core-level spectra to distinguish between haematite (Fe$_{2}$O$_{3}$) and magnetite (Fe$_{3}$O$_{4}$). Elemental distribution and the efficacy of chloride removal were investigated via WDS-EPMA (JEOL JXA-8200) operated at an accelerating voltage of 15~kV to probe the near-surface region modified by the laser treatment.

\section{Results and Discussion}

\subsection{Influence of Repetition Frequency on Surface Morphology}
\begin{figure}[htb]
\centering
\includegraphics[width=0.9\textwidth]{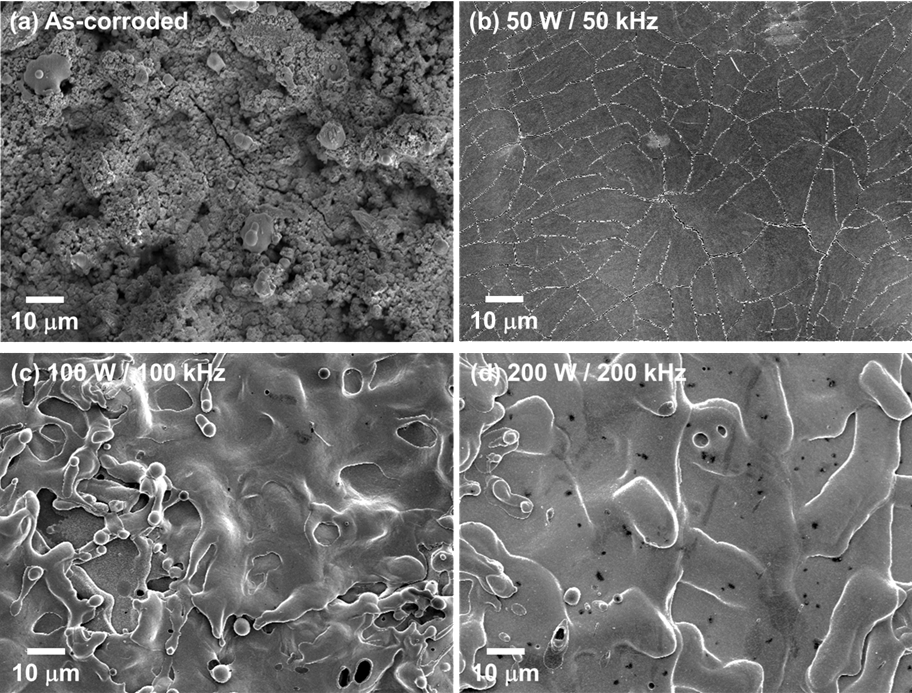}
\caption{SEM micrographs ($V_{\rm acc} = 5$~kV) showing (a) the initial as-corroded surface, alongside laser-treated surfaces under conditions of (b) 50~W / 50~kHz, (c) 100~W / 100~kHz, and (d) 200~W / 200~kHz. Elevating the repetition frequency drives a structural transition from a porous oxide layer to a dense, resolidified surface morphology.}
\label{fig:LC_sem}
\end{figure}

Scanning electron microscopy (SEM) was employed to evaluate the microstructural evolution of the rusted SS400 surfaces as a function of laser repetition frequency. The untreated reference specimen (Fig.~\ref{fig:LC_sem}(a)) exhibited a highly porous, heterogeneous rust layer characteristic of advanced marine atmospheric corrosion \cite{Alcantara2017, Dwivedi2017}. High-magnification observations revealed that this corrosion product consisted primarily of scale-like crystals and fragmented, plate-like particles; such morphologies are typically associated with lepidocrocite ($\gamma$-FeOOH) and goethite ($\alpha$-FeOOH) phases formed under marine exposure \cite{DelaFuente2016}. These porous architectures inherently form fine capillary pathways that retain chloride ions and moisture adjacent to the steel substrate, thereby accelerating subsequent degradation \cite{Melchers2008, Kainuma2009}. Consequently, effective laser processing necessitates modifying the surface morphology to obstruct the ingress of these corrosive agents, either through the complete ablation of the porous capillary networks or via the densification of the rust layer into a less permeable barrier.

Under the 50~W (50~kHz) condition, the porous rust layer was effectively removed, whilst the treated surface as a whole retained a predominantly non-melted morphology (Fig.~\ref{fig:LC_sem}(b)). Although minute micro-cracks were locally observed, these are attributed to the rapid photomechanical and localised thermal stresses induced by laser irradiation. At 50~kHz, the 20~$\mu$s pulse interval allows the thermal energy deposited by the preceding pulse to dissipate significantly prior to the subsequent pulse arrival, thereby minimizing cumulative thermal accumulation. Consequently, this lower-frequency regime facilitates effective decontamination whilst suppressing adverse thermal effects on the steel substrate.

Under the 100~W (100~kHz) and 200~W (200~kHz) conditions, the treated surfaces exhibited a distinct resolidified morphology (Figs.~\ref{fig:LC_sem}(c) and (d)). Although spatial pulse overlap was limited by the 80~$\mu$m pitch, the shortened pulse interval enhanced the likelihood of thermal persistence from the preceding pulse until the subsequent irradiation. Such thermal accumulation occurs when the heat-affected zone induced by a prior pulse fails to dissipate fully within the inter-pulse duration \cite{Siano2010}. Consequently, the progressive thermal accumulation elevates the local temperature sufficiently to induce localized melting at the steel/oxide interface, forming a dense resolidified microstructure via subsequent rapid solidification.

These morphological evolutions directly reflect the progressive intensification of thermal accumulation with increasing repetition frequency. Whilst the substrate morphology remains relatively preserved at 50~kHz, the enhanced heat retention at 100~kHz drives surface densification characterized by localized, melt-resolidified clusters. At 200~kHz, these discrete features amalgamate into a smoother, more continuous resolidified layer, indicating a complete transition in the surface melting behavior. 

To determine whether this structural transition is accompanied by corresponding alterations in the oxide phases, X-ray photoelectron spectroscopy (XPS) was performed. The following analysis specifically verifies the occurrence of a phase transformation from the initially haematite-rich rust layer to a protective, magnetite-rich oxide scale.

\subsection{Surface Chemical State Analysis}
X-ray photoelectron spectroscopy (XPS) was conducted on the surface treated at 100~kHz—where distinctive surface densification was observed via SEM—to evaluate the chemical state of the iron oxides in relation to the morphological evolution. In this study, the 100~kHz condition serves as the representative regime for evaluating the chemical state. Whilst time-resolved waveform measurements at 200~kHz revealed that subsequent pulses arrive prior to the complete decay of the trailing pulse tail, the 100~kHz condition allows each pulse to interact with a more thermodynamically stable transient state, offering a distinct thermal accumulation behavior. 

\begin{figure}[htb]
\centering
\includegraphics[width=0.75\textwidth]{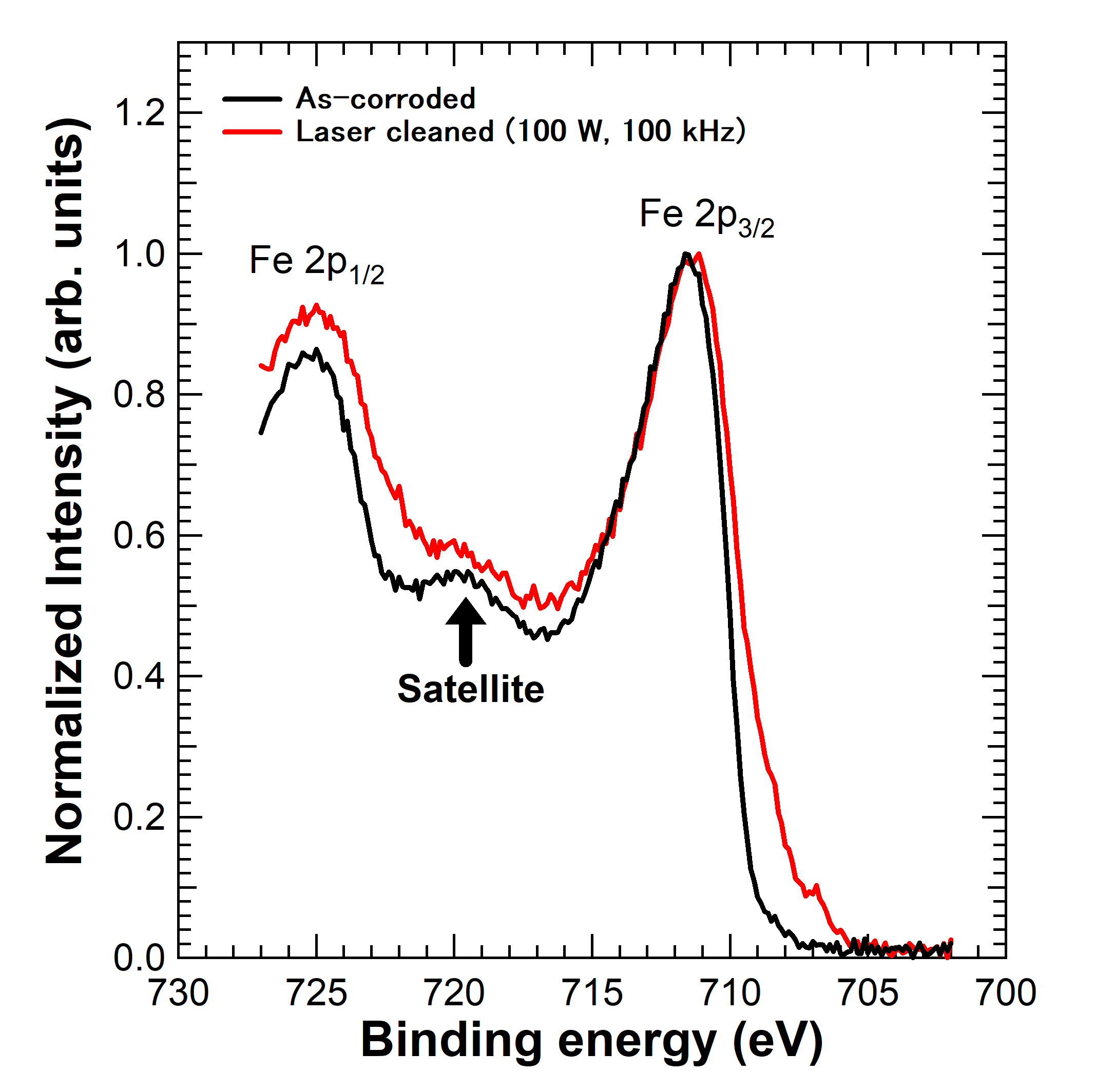}
\caption{High-resolution Fe~2$p$ XPS spectra of the as-corroded surface (red) and the laser-cleaned surface at 100~W (100~kHz) (black). The pronounced satellite peak at approximately 720~eV observed for the as-corroded surface is markedly suppressed after laser cleaning, indicating a change in the surface iron oxide from a Fe$^{3+}$-dominated state toward a magnetite-like mixed-valence state. A rigid binding-energy shift of approximately 1~eV was corrected to compensate for surface charging.}
\label{fig:xps}
\end{figure}

Figure~\ref{fig:xps} compares the intensity-normalised XPS core-level spectra in the Fe~$2p$ region obtained from the untreated reference surface and the surface treated at 100~W (100~kHz). For the untreated specimen, a distinct satellite feature is observed near 720~eV. According to established XPS assignments for iron oxides \cite{Yamashita2008, Yamashita2008_erratum}, this satellite structure is uniquely characteristic of the Fe$^{3+}$ state associated with haematite ($\text{Fe}_{2}\text{O}_{3}$). This observation corroborates the finding that the initial rust layer is predominantly composed of trivalent ferric oxides.

After 100~W laser treatment, the satellite structure near 720~eV is significantly suppressed, whilst a slight broadening is observed in the Fe~$2p_{3/2}$ main peak. These spectral transitions are consistent with a shifting from a chemical state dominated by Fe$^{3+}$ to a mixed-valence Fe$^{2+}$/Fe$^{3+}$ state characteristic of magnetite ($\text{Fe}_{3}\text{O}_{4}$) \cite{Yamashita2008}. In the magnetite phase, the coexistence of Fe$^{2+}$ and Fe$^{3+}$ contributions inherently renders the satellite profile less distinct than that of haematite. Therefore, this reduction in satellite intensity confirms that the observed spectral evolution demonstrates a laser-induced reduction reaction within the surface iron oxide, rather than simply the exposure of the underlying steel substrate.

This interpretation is consistent with stratified rust models in which oxyhydroxides dominate the outer surface whilst magnetite is concentrated within the denser layers adjacent to the steel--rust interface \cite{DelaFuente2016}. In the present process, the pulse-tail heating and frequency-dependent thermal accumulation provide a plausible mechanism for transforming the porous ferric oxide into a dense, magnetite-rich surface layer. The rigid binding-energy shift of approximately 1~eV observed in Fig.~\ref{fig:xps} is attributed to surface charging; this charging state inherently evolves as the highly resistive porous oxide \cite{Dwivedi2017} transforms into a more compact, semiconducting oxide layer.

\subsection{Evaluation of Desalination Efficacy}
WDS-EPMA analysis was conducted to evaluate the elimination of sea-salt-derived residues via the laser treatment. An accelerating voltage of 15~kV was employed to ensure sufficient signal intensity for the characteristic X-rays of Na and Cl. To guarantee the reliability of the analysis, low-magnification measurements were performed over a wide area encompassing multiple surface pits.

\begin{figure}[p]
  \centering
  \includegraphics[height=0.85\textheight, keepaspectratio]{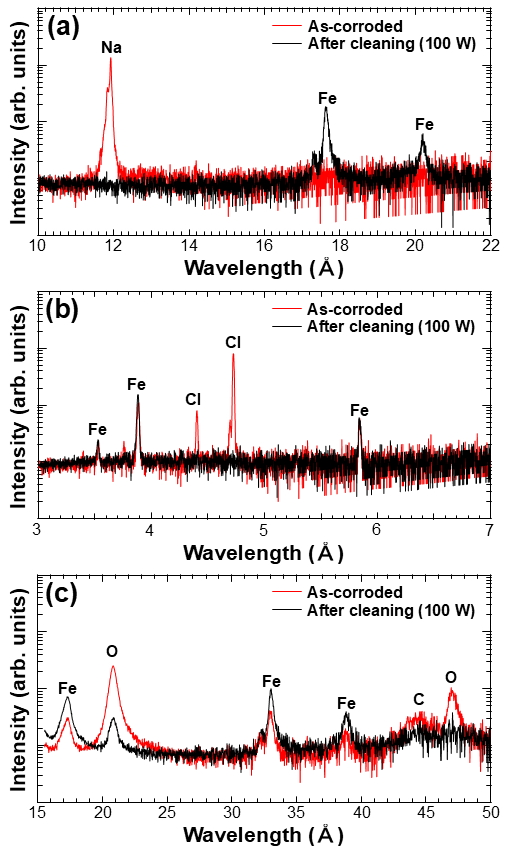}
  
  \caption{High-resolution WDS-EPMA spectra obtained from the untreated reference (red) and the 100~W (100~kHz) laser-treated surface (black), comparing the (a) Na-K$\alpha$, (b) Cl-K$\alpha$, and (c) wide-range O, C, and low-energy Fe regions. The accelerating voltage ($V_{\rm acc} = 15$~kV) is selected to optimize the interaction volume within the laser-modified near-surface region.}
  \label{fig:EPMAfig2}
\end{figure}

Figure~\ref{fig:EPMAfig2} compares the WDS spectra obtained from the untreated reference surface and the surface treated at 100~W (100~kHz). For the untreated specimen, distinct characteristic X-ray peaks corresponding to Na-K$\alpha$ (Fig.~\ref{fig:EPMAfig2}(a)) and Cl-K$\alpha$ (Fig.~\ref{fig:EPMAfig2}(b)) are observed, confirming the presence of sea-salt residues accumulated during the salt spray exposure. Conversely, following the laser treatment, both Na and Cl signals drop to near-background levels, even across the macroscopically rough regions containing multiple surface pits. These results demonstrate the comprehensive removal of sea-salt contaminants, thereby mitigating the risk of subsequent chloride-induced re-corrosion \cite{Bhandari2015, Melchers2008}.

The wide range wavelength dispersive spectra (Fig.~\ref{fig:EPMAfig2}(c)) reveal a significant reduction in both O-K$\alpha$ and carbon-derived signals compared to the untreated reference surface. The decrease in O-K$\alpha$ intensity indicates the elimination of oxygen-rich corrosion products, whilst the attenuation of the carbon signal confirms the removal of surface carbonaceous contaminants.

Synthesizing these findings with the SEM and XPS data establishes that the laser irradiation drives the ablation of the porous, Fe$^{3+}$-rich rust layer, whilst simultaneously transforming the residual surface oxides into a dense, magnetite ($\text{Fe}_{3}\text{O}_{4}$)-dominated state. This physical model is fully consistent with the morphological densification and the suppression of the Fe$^{3+}$ satellite feature detailed above. Furthermore, this structural densification validates the stratified rust model, which characteristically features a dense, protective magnetite scale within the inner region of advanced atmospheric corrosion.

\subsection{Thermal Mechanisms and Phase Transformation}
The collective findings from SEM, XPS, and WDS-EPMA demonstrate that surface modification, chloride removal, and oxide phase transformation proceed concurrently during the laser treatment. To evaluate this phenomenon from a thermodynamic perspective, the relationship between incident laser fluence and transient surface temperature was estimated for the 1064~nm pulsed fibre laser operating at 100~W (100~kHz). Assuming a peak fluence $F_p$ of approximately 20~J/cm$^{2}$, the single-pulse surface temperature rise $\Delta T$ was calculated using a classical one-dimensional heat conduction model:

\begin{equation}
\Delta T \approx \frac{2 \alpha_{\rm eff} F}{\sqrt{\pi \kappa \rho c_p \tau}}
\label{eq:temp_fluence}
\end{equation}

where $F$ is the fluence, $\alpha_{\rm eff}$ is the effective absorptivity of the iron oxide surface at 1064~nm, $\tau$ is the pulse duration, and $\kappa$, $\rho$, and $c_p$ are the thermal conductivity, density, and specific heat capacity of the rust layer, respectively. Given that these thermophysical properties evolve dynamically during ablation and local melting, this analytical solution serves strictly as an order-of-magnitude estimate. Under the high-intensity leading edge of the pulse, the maximum surface temperature briefly reaches the regime ($\approx$3000~K) required for the evaporation or sublimation of iron oxides, providing a sufficient driving force for the explosive ablation of the outer porous rust layer \cite{Zhou2023, Fujita2024_eng}.

Crucially, whilst this analytical model isolates the instantaneous temperature rise induced by an individual pulse, it neglects the cumulative heat retention inherent to high-repetition-rate irradiation. In practice, the 10~$\mu$s pulse interval at 100~kHz forces subsequent pulses to arrive prior to the complete thermal dissipation of preceding pulses, shifting the baseline surface temperature significantly higher than the single-pulse prediction. This progressive thermal accumulation is further augmented by the residual energy delivered during the protracted 600~ns decaying tail of the pulse profile (Fig.~1(b)). Utilizing an effective absorptivity $\alpha_{\rm eff}$ of 0.12 to account for plasma shielding and transient reflectivity drops, the temperature sustained during this trailing edge is estimated to be approximately 1050~K. This localized thermal window is above the threshold required for the haematite-to-magnetite reduction under local redox conditions, whilst remaining safely below the regime for extensive bulk melting of the steel substrate \cite{Spreitzer2019}.

The frequency dependence observed in the SEM images correlates directly with this combined thermal framework. At 50~kHz, the prolonged 20~$\mu$s pulse interval permits substantial cooling between successive pulses, inducing the thermal stress and subsequent micro-cracking detected via SEM. Such sensitivity of microstructural evolution to cooling rates is consistent with established behaviors for laser-treated carbon steels \cite{Bendoumia2020}. Conversely, under high-frequency regimes ($\ge$100~kHz), the shortened inter-pulse duration inhibits complete heat dissipation, maintaining a quasi-plastic or locally molten state for longer and driving the formation of the dense, consolidated morphology. This modified thermal history directly stabilizes the mixed-valence magnetite phase confirmed via XPS, standing in stark contrast to ultra-short picosecond laser processing \cite{Wisse2014} where cold ablation minimizes the heat-affected zone. The nanosecond regime utilized herein thus offers a distinct advantage, acting as an \textit{in situ} laser-induced surface engineering technique \cite{Lahoz2008} that simultaneously ablates corrosion products and consolidates the residual oxide scale into a highly protective surface layer.

\section{Conclusions}
In this study, we demonstrated a single-step, non-contact laser surface modification method for rusted SS400 carbon steel that simultaneously achieves the removal of marine-derived contaminants and the protective phase transformation of the residual oxide scale. By systematically varying the repetition frequency whilst maintaining a constant pulse energy, we elucidated the synergistic effects of the temporal pulse waveform, which features a full width at half maximum of 150~ns and a protracted decay tail extending to 600~ns, on the competitive mechanisms of ablation, thermal accumulation, and metallurgical phase transformation.

WDS-EPMA and Fe~$2p$ XPS analyses demonstrated that treatment at 100~W (100~kHz) effectively removes contaminants such as Na and Cl to near-background levels, even within pits on macroscopically rough surfaces. Simultaneously, it induces a distinct chemical transformation from the initial trivalent ferric rust layer to a dense, mixed-valence magnetite ($\text{Fe}_{3}\text{O}_{4}$) state. This localised phase transformation is fundamentally governed by the waveform characteristics of the laser pulse. Whilst the high peak power at the pulse's leading edge provides the threshold fluence ($\approx$20~J/cm$^{2}$) required to explosively ablate the porous outer rust layer, the subsequent 600~ns trailing tail continuously supplies residual thermal energy to the surface. In the high-frequency range of 100--200~kHz, the shortened pulse interval inhibits complete heat dissipation, resulting in thermal accumulation. This accumulated heat promotes morphological densification and stabilises a highly protective magnetite scale adjacent to the steel substrate, consistent with the stratified rust model.

As a result, this nanosecond-pulse tail heating method has successfully achieved both the removal of chloride contamination and \textit{in situ} surface densification without the use of chemical additives or mechanical abrasives. The resulting dense, magnetite-rich surface layer represents a promising, environmentally friendly approach to mitigating the risk of chloride-induced re-corrosion in structural steel. Whilst the range of appropriate processing conditions has been established, future work must involve rigorous, long-term electrochemical evaluations to directly quantify the corrosion resistance and structural durability of this laser-modified interface under actual outdoor exposure conditions.

\section*{Acknowledgements}
The authors thank the iCeMS Analysis Center at Kyoto University for access to the WDS-EPMA and XPS facilities and for technical support with material characterisation.
This work was supported by the Japan Agency for Medical Research and Development (AMED) under Grant Number JP22am0401020 (Cabinet Office, Government of Japan), the JST FOREST Program (Grant Number JPMJFR203R), the JST-Mirai Program (Grant Number JPMJMI22H5), and JSPS KAKENHI (Grant Number JP23K26777) from the Ministry of Education, Culture, Sports, Science and Technology (MEXT), Japan. Financial support from JKA through the KEIRIN RACE promotion funds, the Inamori Foundation (InaRIS Fellowship Program), and the Asian Young Scientist Fellowship is also gratefully acknowledged.


\begin{thebibliography}{99}



\bibitem{Asmus_OAC}
 L. Lazzarini, L. Marchesini, and J. F. Asmus, Lasers for the cleaning of statuary: Initial results and potentialities, \textit{J. Vac. Sci. Technol.} \textbf{151}  (1973) 1039--1043.

\bibitem{Steen2010}
W. M. Steen and J. Mazumder, \textit{Laser Material Processing}, 4th ed., Springer, London, 2010.

\bibitem{Zhu2022}
G. Zhu, Z. Xu, Y. Jin, X. Chen, L. Yang, J. Xu, D. Shan, Y. Chen, and B. Guo, Mechanism and application of laser cleaning: A review, \textit{Opt. Lasers Eng.} \textbf{157} (2022) 107130.

\bibitem{Bauerle2005}
D. B\"{a}uerle, J. D. Pedarnig, I. Vrejoiu, M. Peruzzi, D. G. Matei and D. Brodoceanu, Laser processing and chemistry: Applications in nanopatterning, material synthesis and biotechnology, \textit{Rom. Rep. Phys.} \textbf{57} (2005) 935--952.

\bibitem{Zapka1993}
W. Zapka, W. Ziemlich, W. P. Leung and A. C. Tam, Laser cleaning: Laser-induced removal of particles from surfaces, \textit{Adv. Mater. Opt. Electron.} \textbf{2} (1993) 63--70.

\bibitem{Tam1992}
A. C. Tam, W. P. Leung, W. Zapka and W. Ziemlich, Laser-cleaning techniques for removal of surface particulates, \textit{J. Appl. Phys.} \textbf{71} (1992) 3515--3523.

\bibitem{Song2002}
W. D. Song, M. H. Hong, H. L. Koh, W. J. Wang, Y. W. Zheng, Y. F. Lu and T. C. Chong, Laser-induced removal of plate-like particles from solid surfaces, \textit{Appl. Surf. Sci.} \textbf{186} (2002) 69--74.

\bibitem{Lu1994}
Y. F. Lu, M. Takai, S. Komuro, T. Shiokawa and Y. Aoyagi, Surface cleaning of metals by pulsed laser irradiation in air, \textit{Appl. Phys. A} \textbf{59} (1994) 281--288.

\bibitem{Zhou2023}
Z. Zhou, W. Sun, J. Wu, H. Chen, F. Zhang and S. Wang, Mechanisms of high-intensity laser cleaning and its applications in industry: a review, \textit{Processes} \textbf{11} (2023) 1445.

\bibitem{Dwivedi2017}
D. Dwivedi, K. Lepkov\'{a} and T. Becker, Carbon steel corrosion: a review of key surface properties and characterization methods, \textit{RSC Adv.} \textbf{7} (2017) 4580--4610.

\bibitem{Zhou2023_Review}
B. Zhou, Y. Li, S. Zhang, G. Zhang and J. Wang, A review of laser cleaning of steel surfaces: Mechanisms, parameters, and industrial applications, \textit{Materials} \textbf{16} (2023) 1420.

\bibitem{Wu2021}
Y. Wu, X. Ren, H. Liu, Y. Yan and X. Kang, Influences of laser parameters on the cleaning quality of carbon steel surface, \textit{Laser Technol.} \textbf{45}(4) (2021) 500--506.

\bibitem{Bhandari2015}
J. Bhandari, F. Khan, R. Abbassi, V. Garaniya and R. Ojeda, Modelling of pitting corrosion in marine and offshore steel structures – A technical review, \textit{J. Loss Prev. Process. Ind.} \textbf{37} (2015) 39--62.

\bibitem{Melchers2008}
R. E. Melchers and R. J. Jeffrey, Probabilistic models for steel corrosion loss and pitting of marine infrastructure, \textit{Reliab. Eng. Syst. Saf.} \textbf{93} (2008) 423--432.

\bibitem{Kainuma2009}
S. Kainuma, N. Hosomi, A. Goto and Y. Itoh, Fundamental study on evaluation for time-dependent corrosion behavior of long steel members in marine environment, \textit{J. JSCE, Ser. A} \textbf{65} (2009) 440--453 (in Japanese).

\bibitem{Alcantara2017}
J. Alc\'{a}ntara, D. de la Fuente, B. Chico, J. Simancas, I. D\'{i}az and M. Morcillo, Marine atmospheric corrosion of carbon steel: A review, \textit{Materials} \textbf{10} (2017) 406.

\bibitem{DelaFuente2016}
D. de la Fuente, J. Alc\'{a}ntara, B. Chico, J. Simancas, I. D\'{i}az and M. Morcillo, Characterisation of rust surfaces formed on mild steel exposed to marine atmospheres using XRD and SEM/Micro-Raman techniques, \textit{Corros. Sci.} \textbf{112} (2016) 530--547.

\bibitem{Fujita2020}
K. Fujita, H. Inagaki, K. Toyosawa, K. Takahara, T. Utsushikawa, H. Fujita, M. Yamada and S. Okihara, kW-class laser cleaning for steel structure maintenance and decontamination, \textit{J. At. Energy Soc. Jpn.} \textbf{62} (2020) 259--262 (in Japanese).


\bibitem{Fujita2024_eng}
K. Fujita, S. Kainuma and H. Shimizu, Potential of Surface Preparation by Continuous Wave Laser Irradiation: Including Efforts for Safety and Human Resource Development by the Laser Construction Research Group, \textit{Proc. 24th Tech. Conf. Japan Association of Steel Bridge and Structure Painting} (2024) (in Japanese).


\bibitem{Siano2010}
S. Siano and R. Salimbeni, Advances in laser cleaning of artwork and objects of historical interest: the optimized pulse duration approach, \textit{Acc. Chem. Res.} \textbf{43} (2010) 739--750.

\bibitem{Ayoola2017}
W. A. Ayoola, W. J. Suder and S. W. Williams, Parameters controlling weld bead profile in conduction laser welding, \textit{J. Mater. Process. Technol.} \textbf{249} (2017) 522--530.

\bibitem{Bendoumia2020}
A. Bendoumia, N. Makuch, R. Chegroune, M. Kulka, M. Keddama, P. Dziarski and D. Przestacki, The effect of temperature distribution and cooling rate on microstructure and microhardness of laser re-melted and laser-borided carbon steels, \textit{Surf. Coat. Technol.} \textbf{387} (2020) 125541.

\bibitem{Choubey2014}
A. Choubey, J. S. Saini, S. K. Rai and K. Rengarajan, Study and development of 22 kW peak power fiber coupled short pulse Nd:YAG laser for cleaning applications, \textit{Opt. Lasers Eng.} \textbf{62} (2014) 69--79.

\bibitem{Ettefagh2020}
A. H. Ettefagh, H. Wen, A. Chaichi, M. I. Islam, F. Lu, M. Gartia and S. Guo, Laser surface modifications of Fe-14Cr ferritic alloy for improved corrosion performance, \textit{Surf. Coat. Technol.} \textbf{381} (2020) 125194.


\bibitem{Wisse2014}
M. Wisse, L. Marot, R. Steiner, D. Mathys, A. Stumpp, M. Joanny, J. M. Travere and E. Meyer, Pico- and nanosecond laser ablation of mixed tungsten / aluminium films, \textit{Fusion Sci. Technol.} \textbf{66} (2014) 308--314.

\bibitem{Lahoz2008}
R. Lahoz, J. P. Espin\'{o}s, G. F. de la Fuente and A. R. Gonz\'{a}lez-Elipe, ``In situ'' XPS studies of laser induced surface cleaning and nitridation of Ti, \textit{Surf. Coat. Technol.} \textbf{202} (2008) 1486--1492.

\bibitem{Grossi2016}
C. M. Grossi and D. Benavente, Colour changes by laser irradiation of reddish building limestones, \textit{Appl. Surf. Sci.} \textbf{384} (2016) 525--529.

\bibitem{Spreitzer2019}
D. Spreitzer and J. Schenk, Iron ore reduction by hydrogen using a laboratory scale fluidized bed reactor: kinetic investigation---experimental setup and method for determination, \textit{Metall. Mater. Trans. B} \textbf{50} (2019) 2471--2484.

\bibitem{Yamashita2008}
T. Yamashita and P. Hayes, Analysis of XPS spectra of $\text{Fe}^{2+}$ and $\text{Fe}^{3+}$ ions in oxide systems, \textit{Appl. Surf. Sci.} \textbf{254} (2008) 2441--2449.

\bibitem{Yamashita2008_erratum}
T. Yamashita and P. Hayes, Erratum to ``Analysis of XPS spectra of $\text{Fe}^{2+}$ and $\text{Fe}^{3+}$ ions in oxide systems'', \textit{Appl. Surf. Sci.} \textbf{255} (2009) 8194.

\end{thebibliography}
\end{document}